\pdfoutput=1
\RequirePackage{ifpdf}
\ifpdf 
\documentclass[pdftex]{sigma}
\else
\documentclass{sigma}
\fi

\def\Z{{\mathbb Z}}
\def\Q{{\mathbb Q}}
\def\F{{\mathbb F}}

\def\P{{\mathbb P}}

\newtheorem{Theorem}{Theorem}[section]

\newtheorem{Proposition}[Theorem]{Proposition}
{\theoremstyle{definition}
\newtheorem{Definition}[Theorem]{Definition}

}

\begin{document}

\allowdisplaybreaks

\renewcommand{\PaperNumber}{056}

\FirstPageHeading

\ShortArticleName{Integrability of Discrete Equations Modulo a~Prime}

\ArticleName{Integrability of Discrete Equations Modulo a~Prime}

\Author{Masataka KANKI}

\AuthorNameForHeading{M.~Kanki}

\Address{Graduate School of Mathematical Sciences,
\\
University of Tokyo, 3-8-1 Komaba, Tokyo 153-8914, Japan}
\Email{\href{kanki@ms.u-tokyo.ac.jp}{kanki@ms.u-tokyo.ac.jp}}

\ArticleDates{Received April 24, 2013, in f\/inal form September 05, 2013; Published online September 08, 2013}

\Abstract{We apply the ``almost good reduction'' (AGR) criterion, which has been introduced in our previous
works, to several classes of discrete integrable equations.
We verify our conjecture that AGR plays the same role for maps of the plane def\/ine over simple f\/inite
f\/ields as the notion of the singularity conf\/inement does.
We f\/irst prove that $q$-discrete analogues of the Painlev\'{e}~III and~IV equations have AGR.
We next prove that the Hietarinta--Viallet equation, a~non-integrable chaotic system also has AGR.}


\Classification{37K10; 34M55; 37P25}

\section{Introduction} The purpose of this paper is to def\/ine the nonlinear discrete integrable equations
over f\/inite f\/ields and to investigate how to formulate the integrability of them over f\/inite
f\/ields, with the help of the theory of arithmetic dynamics.
In the theory of arithmetic dynamics, we are interested in how the properties of the mappings change as we
change the set on which the mappings are def\/ined~\cite{Silverman}.
In particular, the system over the f\/ield of $p$-adic integers and its reduction modulo a~prime to the
f\/inite f\/ield attracts much attention.
We have another interest in the dynamical systems over f\/inite f\/ields in terms of cellular automata, of
which the underlying set consists of a~f\/inite number of elements and the mapping is given by recurrence
formulae~\cite{Wolfram}.

Let us f\/irst explain the problems we encounter and review some of the previous results.
In the case of linear discrete equations, there is no problem in def\/ining the equations over f\/inite
f\/ields just by changing the f\/ield on which the equations are def\/ined to f\/inite f\/ields.
This approach can also be valid for equations such as bilinear equations.
In this direction, the discrete KP and KdV equations and their soliton solutions over f\/inite f\/ields
have been investigated~\cite{Bial}.
However, in order to deal with nonlinear discrete equations, we frequently pass through division by zero
modulo prime and indeterminacies (e.g.~$0/0$, $\infty\pm \infty$), which prevent us from obtaining the evolution.
One of the methods to tackle this dif\/f\/iculty is to restrict the domain of def\/inition so that we do
not need to treat indeterminacies.
For example we can terminate the evolution of the equation when it hits these points.
Integrability of the discrete equations over f\/inite f\/ields has been investigated in terms of the
lengths of the periodic and terminating orbits~\cite{Roberts, RobViv}.
The graph structures of the discrete Toda equation whose dependent variables are limited to non-zero values
have been studied~\cite{Tak}.

We take another approach in this paper: we extend the space of initial conditions instead of restricting it.
We have two main ways of extension.
The f\/irst method is to apply Sakai's theory for discrete Painlev\'{e} equations~\cite{Sakai} over the
f\/inite f\/ield.
According to this theory we can construct the birational mapping over the extended space of initial
conditions by blowing up at each singular point.
The space of initial conditions for the discrete Painlev\'{e} II equation has been established over
$\mathbb{F}_p$ and the bijection between a~f\/inite number of points has been constructed~\cite{KMTT2}.
The second approach uses the f\/ield of $p$-adic numbers.
We study this method in detail from here on.
We def\/ine the discrete integrable equations over the f\/ield of $p$-adic numbers $\mathbb{Q}_p$, and then
def\/ine them over $\mathbb{F}_p$, so that they are compatible with the reduction modulo prime from those
over $\mathbb{Q}_p$.
Rational mappings are said to have good reduction if, roughly speaking, the reduction and the evolution of
the system commute.
One of the typical examples with good reduction is the fractional linear transformation related to the
projective linear group PGL$_2$.
Recently, birational mappings over f\/inite f\/ields have been investigated in terms of
integrability~\cite{Roberts}.
In the previous papers, we def\/ined the generalized notion of good reduction so that it could be applied
to wider class of integrable mappings.
We called this notion ``almost good reduction'' (AGR), and proved that discrete and $q$-discrete Painlev\'{e}
II equations have AGR~\cite{KMTT, KMTT2}.
Our conjecture was that AGR is also satisf\/ied for other discrete Painlev\'{e} equations and that AGR is
closely related to the integrability of dynamical systems over f\/inite f\/ields.
In this paper, we prove that several types of $q$-discrete analogues of the Painlev\'{e}
equations~\cite{RGH} have AGR for an appropriate domain, thereby verifying the conjecture.
We also study the application of AGR to a~chaotic system~-- Hietarinta--Viallet equation~\cite{HV}~-- and
conclude that AGR can be seen as an arithmetic analogue of the singularity conf\/inement
test~\cite{Grammaticosetal}.

\section{Reduction modulo a~prime}

Let $p$ be a~prime number.
Each non-zero rational number $x \in \Q^{\times}$ has a~unique representation $x=p^{v_p(x)}\frac{u}{v}$
where $v_p(x), u, v \in \Z$ and $u$ and $v$ are coprime integers neither of which is divisible by~$p$.
The $p$-adic norm $|x|_p$ is def\/ined as $|x|_p:=p^{-v_p(x)}$
($|0|_p:=0$). The f\/ield of $p$-adic numbers~$\Q_p$ is def\/ined as a~completion of $\Q$ with respect to
the $p$-adic norm.
The ring of $p$-adic integers is def\/ined as
\begin{gather*}
\Z_p:=\{x\in\Q_p\, |\,|x|_p\le1\}.
\end{gather*}
The ring $\Z_p$ has the unique maximal ideal
\begin{gather*}
\mathfrak{p}=p\Z_p=\{x\in\Z_p\, |\, |x|_p<1\}.
\end{gather*}
We def\/ine the reduction of $x$ modulo $\mathfrak{p}$ by
\begin{gather*}
\Z_p\ni x\mapsto(x\!\mod\mathfrak{p})\in\Z_p/\mathfrak{p}\cong\F_p,
\end{gather*}
and denote $(x\! \mod \mathfrak{p})$ as $\tilde{x}$.
The above mapping def\/ines a~reduction of $p$-adic integers to the (simple) f\/inite f\/ield.
Note that if we limit $x$ to be a~(rational) integer, then $\tilde{x}$ is nothing but $x$ modulo $p$.
The reduction is naturally generalised to $\Q_p$:
\begin{gather*}
\Q_p\ni x\mapsto
\begin{cases}\tilde{x}, & x\in\Z_p,
\\
\infty, & x\in\Q_p\setminus\Z_p
\end{cases}
\in\P\F_p.
\end{gather*}
We also denote the right hand side by $\tilde{x}$.
Here the space $\P\F_p$ denotes the projective line $\P^1 (\F_p)$ def\/ined over the f\/inite f\/ield
$\F_p$.
As a~set, we have $\P\F_p=\F_p\cup\{\infty\}$.

Next we def\/ine the reduction of maps of the plane.
Let $\phi$ be a~rational map of the plane $\Q_p^2$ given by two rational functions def\/ined over
$(x,y)\in\mathbb{Q}_p^2$:
\begin{gather*}
\phi(x,y)=(f(x,y),g(x,y)),
\end{gather*}
where $f,g\in\mathbb{Q}_p(x,y)$ are rational functions.
By multiplying numerators and denominators of $f$ and $g$ by suitable powers of $p$, the coef\/f\/icients
of both $f$ and $g$ can be taken in $\Z_p$ and that at least one of the coef\/f\/icients is in
$\Z_p^{\times}$.
From here on we assume this ``minimal'' form for rational functions.        
If neither of the denominator of the minimal form of $f$ or that of $g$ modulo $\mathfrak{p}$ is zero, then
$\tilde{\phi}$ is def\/ined as the system whose coef\/f\/icients are reduced to $\mathbb{F}_p$:
\begin{gather*}
\tilde{\phi}(x,y)=\big(\tilde{f}(x,y),\tilde{g}(x,y)\big) \in(\mathbb{F}_p(x,y))^2,
\end{gather*}
where $(x,y)\in\Q_p^2$.
The map $\phi$ is said to have \textit{good reduction} (modulo $\mathfrak{p}$) on the domain
$\mathcal{D}\subseteq\mathbb{Z}_p^2$, if we have
\begin{gather*}
\widetilde{\phi(x,y)}=\tilde{\phi}(\tilde{x},\tilde{y})
\end{gather*}
for any $(x,y) \in \mathcal{D}$~\cite{Silverman}.
Although the good reduction is useful in arithmetic dynamical systems, the discrete Painlev\'{e} equations
(expressed as a~dynamical system) do not have good reduction since they frequently pass through
singularities after reducing the equations modulo a~prime.
Also, the discrete Painlev\'{e} equations are non-autonomous mappings, and we need a~generalization of good
reduction to a~non-autonomous mapping.
Therefore, we have modif\/ied the good reduction so that it can be applied to wider class of systems, in
particular to the discrete Painlev\'{e} equations.

\begin{Definition}[\cite{KMTT}] A (non-autonomous) rational map of the plane $\phi_n$ has almost good reduction (AGR) modulo
$\mathfrak{p}$ on the domain $\mathcal{D}^{(n)}\subseteq\Z_p^2\cap\phi_n^{-1}(\mathbb{Q}_p^2)$ if for any
$\text{p}=(x,y) \in \mathcal{D}^{(n)}$ and any time step $n$, there exists a~positive integer
$m_{\text{p};n}$ such that
\begin{gather*}
\widetilde{\phi_n^{m_{\text{p};n}}(x,y)}
=\widetilde{\phi_n^{m_{\text{p};n}}}(\tilde{x},\tilde{y}).
\end{gather*}
Here, the iteration $\phi_n^m$ is def\/ined as
$\phi_n^m:=\left(\phi_{n+m-1}\circ\phi_{n+m-2}\circ\dots\circ\phi_n\right)|_{\mathcal{D}^{(n)}}$ for $m>0$.
\end{Definition}

If the domain $\mathcal{D}^{(n)}$ does not depend on $n$, we just denote it by $\mathcal{D}$.
Note that, in particular, if we can take $m_{\text{p};n}=1$ for all points $\text{p} \in
\mathcal{D}^{(n)}$ and all $n$, then the mapping $\phi_n$ has good reduction.
Therefore AGR is weaker than good reduction.

The following simple mapping $\Psi_\gamma$ illustrates how almost good reduction works.
Let us def\/ine
\begin{gather}
\Psi_\gamma: \
\begin{cases}
x_{n+1} =\dfrac{ax_n+1}{x_n^\gamma y_n},
\\
y_{n+1} =x_n,
\end{cases}
\label{discretemap}
\end{gather}
where $|a|_p\le1$ and $\gamma \in \Z_{\ge 0}$ are parameters.
Note that we omitted the cases of $|a|_p>1$, since we have $v_p(a)<0$ in this case, and we have to deal
with a~mapping such as
\begin{gather*}
\phi:\ x_{n+1}=\frac{\frac{x_n}{p}+1}{x_n^\gamma y_n}=\frac{x_n+p}{p x_n^\gamma y_n},
\qquad
y_{n+1}=x_n,
\end{gather*}
whose reduction of coef\/f\/icients $\widetilde{\phi}$ is not well-def\/ined (note that
$\widetilde{1/p}=\infty$).
The map~\eqref{discretemap} is known to be integrable if and only if $\gamma=0,1,2$.
In these cases the map is a~symmetric QRT mapping~\cite{QRT}.
We have proved in our previous work that the following proposition holds.

\begin{Proposition}[\cite{KMTT}]\label{PropQRT}
The rational mapping~\eqref{discretemap} has almost good reduction modulo $\mathfrak{p}$ on
the domain $\mathcal{D}$ if and only if $\gamma=0,1,2$.
Here $\mathcal{D}:=\Z_p^2\cap \Psi_{\gamma}^{-1}(\mathbb{Q}_p^2)$.
If $\gamma= 1,2$ then $\mathcal{D}=\big\{(x,y) \in \Z_p^2 \,|\, x \ne 0, \; y \ne 0\big\}$.
If $\gamma=0$ then $\mathcal{D}=\big\{(x,y) \in \Z_p^2 \, |\, y \ne 0\big\}$.
\end{Proposition}
We have also proved that the discrete and $q$-discrete Painlev\'{e} II equations also have almost good
reduction~\cite{KMTT, KMTT2}.
From these observations we have conjectured that, for the map of the plane $\Phi_n$ def\/ined over the
f\/ield of $p$-adic numbers, having almost good reduction on the domain $\mathcal{D}^{(n)}=\Z_p^2\cap
\Phi_n^{-1}(\mathbb{Q}_p^2)$ is equivalent to passing the singularity conf\/inement
test~\cite{Grammaticosetal}.
In this article, we support this conjecture by presenting further applications of the almost good reduction
principle to other integrable equations such as several types of $q$-discrete Painlev\'{e} equations and
a~chaotic equation.

Note that in this paper we only deal with simple f\/inite f\/ields $\F_p$.
To study the equations over a~general f\/inite f\/ield $\F_{p^m}$ $(m>1)$, we need to use the f\/ield
extension of $\Q_p$.
Since a~f\/ield extension $L$ of f\/inite degree $m$ over $\Q_p$ is a~simple extension, there exists an
element $\alpha\in L$ such that $L=\Q_p(\alpha)$.
The reduction from $L$ to the extension of the f\/inite f\/ield $\mathbb{F}_p(\alpha)$ is def\/ined as
\begin{gather*}
L\ni\sum_{i=0}^{m-1}x_i\alpha^i\mapsto
\begin{cases}\displaystyle \sum_{i=0}^{m-1}\tilde{x}_i\alpha^i,& \forall\, i \ x_i\in\Z_p ,
\\
\infty, & \exists \, i \ x_i\in\Q_p\setminus\Z_p .
\end{cases}
\end{gather*}
From $\F_p(\alpha)\cong \F_{p^m}$, we obtain a~system over a~general f\/inite f\/ield.
Since the $p$-adic norm of $\Q_p$ is extended to $L$, propositions in this paper are still valid for $L$,
with slight modif\/ications (e.g.~$\mathcal{D}$ becomes $\left(\Z_p^{\oplus m}\right)^2\cap \Psi_{\gamma}^{-1}(L^2)$ in
Proposition~\ref{PropQRT}).

\section[$q$-difference analogue of Painlev\'{e} equations over a finite field]{$\boldsymbol{q}$-dif\/ference analogue of Painlev\'{e} equations over a~f\/inite f\/ield}

In this section we
prove that the $q$-discrete analogues of Painlev\'{e} III and IV equations have almost good reduction.
These two equations are indeed integrable in the sense that they pass the singularity conf\/inement
test~\cite{Grammaticosetal}.
Note that, although passing singularity conf\/inement test is not equivalent to the integrability of the
given discrete equation, singularity conf\/inement can be seen as a~discrete analogue of the Painlev\'{e}
property of the continuous Painlev\'{e} equations.
In the geometrical setting, these two equations have rational surfaces of initial conditions on which the
equations are birational~\cite{Sakai}.
From here on we occasionally write $a\equiv b$ for $a,b\in\mathbb{Z}_p$, to indicate that
$\tilde{a}=\tilde{b}\in\F_p$.

\subsection[$q$-discrete Painlev\'{e} III equation]{$\boldsymbol{q}$-discrete Painlev\'{e} III equation}

The $q$-discrete analogue of Painlev\'{e} III equation
has the following form
\begin{gather*}
x_{n+1}x_{n-1}=\frac{ab(x_n-cq^n)(x_n-dq^n)}{(x_n-a)(x_n-b)},
\end{gather*}
where $a$, $b$, $c$, $d$ and $q$ are parameters~\cite{RGH}.
It is convenient to rewrite it as the following coupled system form
\begin{gather}
\Phi_n: \
\begin{cases}
x_{n+1} =\dfrac{ab\big(x_n-cq^n\big)\big(x_n-dq^n\big)}{y_n(x_n-a)(x_n-b)},
\\
y_{n+1} =x_n.
\end{cases}
\label{qP3}
\end{gather}
\begin{Proposition}\label{PropqP3}
Suppose that $a$, $b$, $c$, $d$, $q$ are distinct parameters with $|a|_p=|b|_p=|c|_p=|d|_p=1$, and we also suppose that
$a+b\not\equiv (c+d)q^3$ and $a\not\equiv b$, then the mapping~\eqref{qP3} has almost good reduction modulo
$\mathfrak{p}$ on the domain $\mathcal{D}:=\big\{(x,y)\in \Z_p^2\,|\, x\neq a,b,\,y\neq 0 \big\}$.
\end{Proposition}
\begin{proof}
Let $(x_{n+1},y_{n+1})=\Phi_n(x_n,y_n)$.
In the case when $\tilde{x}_n\neq \tilde{a},\tilde{b}$ and $\tilde{y}_n\neq 0$, we have
\begin{gather*}
\tilde{x}_{n+1} =\dfrac{\tilde{a}\tilde{b}\big(\tilde{x}_n-\tilde{c}\tilde{q}^n\big)
\big(\tilde{x}_n-\tilde{d}\tilde{q}^n\big)}{\tilde{y}_n(\tilde{x}_n-\tilde{a})(\tilde{x}_n-\tilde{b})},
\\
\tilde{y}_{n+1} =\tilde{x}_n.
\end{gather*}
since the reduction modulo $p\Z_p$ is a~ring homomorphism.
Hence clearly $\widetilde{\Phi_n(x_n,y_n)}=\widetilde{\Phi_n}(\tilde{x}_n,\tilde{y}_n)$.
Next we examine other cases.
We investigate the iterated maps for every initial condition $(x_n,y_n)\in\mathcal{D}$.
There are six cases to consider.
They are essential since the behaviors around the singular points are involved.
From here we sometimes abbreviate $\tilde{a}$ as $a$, $\tilde{b}$ as $b$ for simplicity.

(i) Let us f\/irst consider the case where
\begin{gather*}
x_n\equiv a \qquad \mbox{and}\qquad (a-b)(a+b-cq-dq)\tilde{y}_n\not\equiv b(a-c)(a-d).
\end{gather*}
In this case, $\widetilde{\Phi_n}(\tilde{a},\tilde{y}_n)$ is not well-def\/ined.
This is because we have $\tilde{x}_n-\tilde{a}=\tilde{a}-\tilde{a}=0$ in the denominator of
$\tilde{x}_{n+1}$.
We also learn from $y_{n+2}=x_{n+1}$, that $\tilde{y}_{n+2}$ is not def\/ined either.
Therefore $\widetilde{\Phi_n^2}(\tilde{a},\tilde{y}_n)$ is not well-def\/ined.
Here $\widetilde{\Phi_n^2}:=\widetilde{\Phi_{n+1}\circ \Phi_n}$.

However, at the third iteration step, $\widetilde{\Phi_n^3}(\tilde{a},\tilde{y}_n)$ is well-def\/ined and
we have
\begin{gather*}
\widetilde{\Phi_n^3(x_n,y_n)}=\widetilde{\Phi_n^3}(\tilde{x}_n=\tilde{a},\tilde{y}_n)
=\left(\frac{a(b-cq^2)(b-dq^2)\tilde{y}_n}{b(a-c)(a-d)-(a-b)(a+b-cq-dq)\tilde{y}_n},b\right).
\end{gather*}
From the assumption of the case (i), the denominator is nonzero and is in $\F_p^{\times}$.

(ii) Next we investigate the case of
\begin{gather*}
x_n\equiv a \qquad \mbox{and} \qquad (a-b)(a+b-cq-dq)\tilde{y}_n\equiv b(a-c)(a-d)
\end{gather*}
In this case, none of $\widetilde{\Phi_n^i}(\tilde{a},\tilde{y}_n)$ is well-def\/ined for $i=1,2,3,4$.
Next we calculate the f\/ifth iteration $\Phi_n^5$ at $\tilde{y}_n\equiv
\frac{b(a-c)(a-d)}{(a-b)(a+b-cq-dq)}$ and simplify the outcome.
Then we learn that $\widetilde{\Phi_n^5}(\tilde{a},\tilde{y}_n)$ is well-def\/ined and we have
\begin{gather*}
\widetilde{\Phi_n^5(x_n,y_n)}=\widetilde{\Phi_n^5}(\tilde{x}_n=\tilde{a},\tilde{y}
_n)=\left(\frac{b(a-cq^4)(a-dq^4)}{(a-b)(a+b-cq^3-dq^3)},a\right).
\end{gather*}

(iii) If $\tilde{x}_n=\tilde{b}$ and $(a-b)(a+b-cq-dq)\tilde{y}_n \not \equiv -a(b-c)(b-d)$, by an argument
similar to that in (i), we have
\begin{gather*}
\widetilde{\Phi_n^3(x_n,y_n)}=\widetilde{\Phi_n^3}(\tilde{x}_n=\tilde{b},\tilde{y}_n)
=\left(\frac{b(a-cq^2)(a-dq^2)\tilde{y}_n}{a(b-c)(b-d)+(a-b)(a+b-cq-dq)\tilde{y}_n},a\right).
\end{gather*}

(iv) If $\tilde{x}_n=\tilde{b}$ and $(a-b)(a+b-cq-dq)\tilde{y}_n \equiv -a(b-c)(b-d)$, by an argument
similar to that in (ii), we have
\begin{gather*}
\widetilde{\Phi_n^5(x_n,y_n)}=\widetilde{\Phi_n^5}(\tilde{x}_n=\tilde{b},\tilde{y}
_n)=\left(-\frac{a(b-cq^4)(b-dq^4)}{(a-b)(a+b-cq^3-dq^3)},b\right).
\end{gather*}

(v) If $\tilde{y}_n=0$ and $\tilde{x}_n\not = 0$,
\begin{gather*}
\widetilde{\Phi_n^3(x_n,y_n)}=\widetilde{\Phi_n^3}(\tilde{x}_n,\tilde{y}_n=0)=\left(0,\frac{ab}{\tilde{x}_n}
\right).
\end{gather*}

(vi) If $\tilde{y}_n=0$ and $\tilde{x}_n = 0$,
\begin{gather*}
\widetilde{\Phi_n^4(x_n,y_n)}=\widetilde{\Phi_n^4}(\tilde{x}_n=0,\tilde{y}_n=0)=\left(0,0\right).
\end{gather*}
We have now fully investigated the behaviours around singularities and have completed the proof.
\end{proof}

\subsection[$q$-discrete Painlev\'{e} IV equation]{$\boldsymbol{q}$-discrete Painlev\'{e} IV equation}

The $q$-discrete analogue of Painlev\'{e} IV equation
has the following form
\begin{gather*}
(x_{n+1}x_n-1)(x_nx_{n-1}-1)=\frac{aq^{2n}(x_n^2+1)+bq^{2n}x_n}{cx_n+dq^n},
\end{gather*}
where $a$, $b$, $c$, $d$ and $q$ are parameters~\cite{RG, RGH}.
It can be rewritten as follows:
\begin{gather}
\Phi_n: \
\begin{cases}
x_{n+1} =\dfrac{\tau^2(ax_n^2+bx_n+a)+(x_ny_n-1)(x_n+\tau)}{x_n(x_ny_n-1)(x_n+\tau)},
\\
y_{n+1} =x_n,
\end{cases}
\label{qP4}
\end{gather}
where $\tau=q^n\tau_0$.
Here we took $\tau_0=d/c$ and redef\/ined $a$, $b$ as $ac/d^2 \to a$ and $bc/d^2 \to b$.
\begin{Proposition}
Suppose that $|a|_p=|b|_p=|q|_p=|\tau_0|_p=1$, and we also suppose that $aq^2\tau_0\not \equiv 1$ and
$aq^4\tau_0\not \equiv 1$.
Then the mapping~\eqref{qP4} has almost good reduction modulo $\mathfrak{p}$ on the domain
$\mathcal{D}^{(n)}:=\big\{(x,y)\in \Z_p^2\,\big|\, x\neq 0,\;xy\neq 1,\;x\neq -q^n\tau_0\big\}$.
\end{Proposition}

\begin{proof}
In the proof we use the abbreviation as $\tilde{a}\to a,\;\tilde{b}\to b,\tilde{\tau}_0\to\tau_0$.
By an argument similar to that in Proposition~\ref{PropqP3}, we have only to consider the cases at the
singular points modulo a~prime.

(i) If $\tilde{x}_n=0$ and $1+q^2(-1+a\tau_0-b\tau_0^2+q\tau_0^2+\tau_0 y_n-a\tau_0^2 y_n)\not\equiv 0$,
the f\/irst and second iterations, $\widetilde{\Phi_n}(0,\tilde{y}_n)$ and
$\widetilde{\Phi_n^2}(0,\tilde{y}_n)$ are not well-def\/ined.
However, at the third iteration we have
\begin{gather*}
\widetilde{\Phi_n^3(x_n,y_n)}=\widetilde{\Phi_n^3}(\tilde{x}_n=0,\tilde{y}_n)
\\
\phantom{\widetilde{\Phi_n^3(x_n,y_n)}}
=\left(\frac{-1-q^3\tau_0^2-bq^4\tau_0^2+aq^6\tau_0^3+q^2(1+b\tau_0^2-\tau_0\tilde{y}_n+a\tau_0^2\tilde{y}_n)}
{q^2\tau_0\{1+q^2(-1+a\tau_0-b\tau_0^2+q\tau_0^2+\tau_0\tilde{y}_n-a\tau_0^2\tilde{y}_n)\}}
,-q^2\tau_0\right).
\end{gather*}

(ii) If $\tilde{x}_n=0$ and $1+q^2(-1+a\tau_0-b\tau_0^2+q\tau_0^2+\tau_0 y_n-a\tau_0^2 y_n)\equiv 0$, we
iterate the map further from (i) until the reduced map is well-def\/ined at
$\tilde{y}_n=\frac{1+q^2(-1+a\tau_0-b\tau_0^2+q\tau_0^2)}{q^2\tau_0 (a\tau_0-1)}$.
At the f\/ifth iteration,
\begin{gather*}
\widetilde{\Phi_n^5(x_n,y_n)}=\widetilde{\Phi_n^5}(\tilde{x}_n=0,\tilde{y}
_n)=\left(\frac{-1+q^2+aq^4\tau_0+q^7\tau_0^2-bq^8\tau_0^2}{q^4\tau_0(-1+aq^4\tau_0)},0\right).
\end{gather*}
Since we assumed that $aq^4\tau_0\not\equiv 1$, it is well-def\/ined.

(iii) If $\tilde{x}_n=-q^n\tau_0$ and $\tilde{y}_n\neq -\tau_0^{-1}$,
\begin{gather*}
\widetilde{\Phi_n^3(x_n,y_n)}=\widetilde{\Phi_n^3}(\tilde{x}_n=-q^n\tau_0,\tilde{y}_n)
\\
\qquad
=\left(\frac{-1-\tau_0\tilde{y}_n+\big(q^3-bq^4\big)\tau_0^2(1+\tau_0\tilde{y}
_n)+q^2\{1+b\tau_0^2+\tau_0\tilde{y}_n+a\tau_0^2(-\tau_0+\tilde{y}_n)\}}
{q^2\tau_0(-1+aq^2\tau_0)(1+\tau_0\tilde{y}_n)},0\right),
\end{gather*}
where we assumed $aq^2\tau_0\not\equiv 1$.

(iv) If $\tilde{x}_n=-q^n\tau_0$ and $\tilde{y}_n= -\tau_0^{-1}$,
\begin{gather*}
\widetilde{\Phi_n^5(x_n,y_n)}=\widetilde{\Phi_n^5}\big(\tilde{x}_n=-q^n\tau_0,\tilde{y}_n=-\tau_0^{-1}
\big)=\left(-\frac{1}{aq^6\tau_0^2},-aq^6\tau_0^2\right).
\end{gather*}

(v) If $\tilde{x}_n\tilde{y}_n=1$,
\begin{gather*}
\widetilde{\Phi_n^5(x_n,y_n)}=\widetilde{\Phi_n^5}\left(\tilde{x}_n=\frac{1}{\tilde{y}_n},\tilde{y}
_n\right)=\left(\frac{1}{aq^6\tau_0^3\tilde{y}_n},aq^6\tau_0^3\tilde{y}_n\right).\tag*{\qed}
\end{gather*}
\renewcommand{\qed}{}
\end{proof}

\section[Hietarinta-Viallet equation over a finite field]{Hietarinta--Viallet equation over a~f\/inite f\/ield}

 The Hietarinta--Viallet equation~\cite{HV}
is the following dif\/ference equation:
\begin{gather}\label{HV1}
x_{n+1}+x_{n-1}=x_n+\frac{a}{x_n^2},
\end{gather}
with $a$ as a~parameter.
The equation~\eqref{HV1} passes the singularity conf\/inement test~\cite{Grammaticosetal}, which is
a~notable test for integrability of equations, yet is \textit{not} integrable in the sense that its
algebraic entropy~\cite{BV} is positive and that the orbits display chaotic behaviour~\cite{HV,Takenawa}.
We prove that the AGR is satisf\/ied for this Hietarinta--Viallet equation.
We rewrite~\eqref{HV1} as the following coupled system:
\begin{gather}
\Phi_n: \
\begin{cases}
 x_{n+1} =x_n+\dfrac{a}{x_n^2}-y_n,
\\
y_{n+1} =x_n.
\end{cases}
\label{HV}
\end{gather}

\begin{Proposition}
Suppose that $|a|_p=1$, then the mapping~\eqref{HV} has almost good reduction modulo~$\mathfrak{p}$ on the
domain $\mathcal{D}:=\big\{(x,y)\in \Z_p^2\,|\,x\neq 0\big\}$.
\end{Proposition}

\begin{proof}
If $\tilde{x}_n\neq 0$ then, the next step is immediately well-def\/ined:
$\widetilde{\Phi_n(x_n,y_n)}=\widetilde{\Phi_n}(\tilde{x},\tilde{y})$.

If $\tilde{x}_n=0$, we have to iterate the map four times to obtain
\begin{gather*}
\widetilde{\Phi_n^4(x_n,y_n)}=\widetilde{\Phi_n^4}(\tilde{x}_n=0,\tilde{y}_n)=(\tilde{y}_n,0).\tag*{\qed}
\end{gather*}\renewcommand{\qed}{}
\end{proof}

Therefore we learn that the AGR works similarly to the singularity conf\/inement test in distinguishing the
integrable systems from the non-integrable ones.
In fact, the AGR can be seen as an arithmetic analogue of the singularity conf\/inement test.

\section{Concluding remarks}

We studied the integrable discrete equations over a~f\/inite f\/ield by
reducing them from the f\/ield of $p$-adic numbers.
We considered the ``almost good reduction'' (AGR), which had been proposed to be closely related to the
integrability of discrete dynamical systems over f\/inite f\/ields.
We proved that $q$-discrete Painlev\'{e}~III and IV equations also have AGR, which has been a~conjecture in
our previous article.
We also treated the Hietarinta--Viallet equation, which is non-integrable yet passes singularity
conf\/inement test.
We proved that it also has AGR.
From these observations, we have concluded that AGR is not a~complete integrability test, however, we can
safely state that the AGR is an arithmetic dynamical analogue of the singularity conf\/inement method.
One of the future problems is to modify AGR so that it can identify the systems which are non-integrable
yet pass the singularity conf\/inement test, like the Hietarinta--Viallet equation.
Other future problems are listed below: (i)~to study the geometric construction of the initial value space
by blowups of the projective space of $\Q_p$, and its relation to the Sakai's theory~\cite{Sakai}, (ii)~to
study the relation of our methods to the algebraic entropy~\cite{BV} and its arithmetic
analogue~\cite{Halburd}, (iii)~to formulate the properties of the reduction modulo prime of the higher
dimensional mappings like those in~\cite{KNY}, (iv)~to extend our methods to lattice equations with soliton
solutions, such as the discrete Korteweg--de~Vries equation and the discrete nonlinear Schr\"{o}dinger
equation.

\subsection*{Acknowledgements}

The author wish to thank Professors Jun Mada, K.M.~Tamizhmani, Tetsuji
Tokihiro and Ralph Willox for insightful discussions and comments.
He also thanks the detailed suggestions by the referees.
This work is supported by Grant-in-Aid for JSPS Fellows (24-1379).

\pdfbookmark[1]{References}{ref}
\LastPageEnding

\end{document}